\documentstyle[aps,pra,epsfig,twocolumn]{revtex}

\def\be{\begin{equation}}
\def\ee{\end{equation}}
\def\bea{\begin{eqnarray}}
\def\eea{\end{eqnarray}}
\def\bma{\begin{mathletters}}
\def\ema{\end{mathletters}}
\def\C{\hbox{$\mit I$\kern-.7em$\mit C$}}
\newcommand{\one}{\mbox{$1 \hspace{-1.0mm}  {\bf l}$}}

\begin{document}
\draft

\title{Irreversibility in asymptotic manipulations of entanglement}

\author{G. Vidal and J. I. Cirac}

\address{Institut f\"ur Theoretische Physik, Universit\"at Innsbruck,
A-6020 Innsbruck, Austria}

\date{\today}

\maketitle

\begin{abstract}
We show that the process of entanglement distillation is
irreversible by showing that the entanglement cost of a bound
entangled state is finite. Such irreversibility remains even if extra pure entanglement is loaned to assist the distillation process.
\end{abstract}


\narrowtext

The appearance of irreversibility in physical processes can be
regarded as one of the most fundamental and studied problems in
the history of Physics. In the context of Quantum Information,
it has been pointed out that an irreversible loss of
entanglement might be present in the process of entanglement
distillation \cite{Be960,Be93}. That is, the amount of pure
entanglement that can be distilled out of $N$ copies of some
state $\rho$ might be strictly smaller than the one needed to
create those copies if only local operations and classical
communication (LOCC) are allowed and in the asymptotic limit
($N\to\infty$). Although when $\rho$ represents a pure state
this process is known to be reversible \cite{Be96}, it is
generally believed that for mixed states this is not the case
\cite{note1}. This last statement has not been proved so far
\cite{Ho00}. In this Letter we prove it, i.e. we show, by means of an example, that the
process of entanglement distillation is inherently irreversible. We will also extend this result to a broader context set by catalytic local operations and classical communication (LOCCc) \cite{JP}, where pure entanglement ---to be subsequently returned--- is loaned to assist the distillation process.

Perhaps, the strongest indication that we have so far of the
irreversibility of entanglement distillation is given by the
existence of so--called bound entangled states \cite{Ho98}.
Those are states from which no entanglement can be distilled but
for which, in order to create a single copy, entanglement is
required. Notice that, in spite of being very suggestive, this indication is not conclusive.
It does not rule out the possibility that, in order to
create a larger number of copies, the amount of entanglement
needed per copy vanishes in the asymptotic limit. Although this
seems unlikely, it has not been disproved so far. On the other hand, it 
is also not clear yet whether bound entangled states can be activated, 
and ultimately distilled, with the help of some borrowed pure entanglement. This would still leave an open door for some form of catalytic reversibility.

In the present 
work we will show that the bound entangled state with positive partial transposition (PPT) introduced in Ref.\ \cite{Be99} has a non--vanishing entanglement cost in the asymptotic limit. We will also show that no more pure entanglement can be distilled from PPT states, than just the amount that may have been used in order to assist the distillation process.
In this way, the irreversibility of the
asymptotic manipulation of entanglement in the context of LOCC ---and also in that of LOCCc--- will immediately follow.

Let us formulate more precisely the problem. We consider two
parties located in spatially separated regions and possessing
$N$ copies of the state $\rho$. Let us consider a transformation
$\rho^{\otimes N}\to \rho{\prime}_N$ which fulfills
\be
\lim_{N\to\infty} D(\rho^{\prime}_N,|\Psi\rangle\langle
\Psi|^{\otimes M}) \to 0,
\ee
for some integer $M$ depending on $N$, where $|\Psi\rangle\equiv
(|0,1\rangle-|1,0\rangle)/\sqrt{2}$ is the two--qubit singlet
state and $D$ is a properly chosen distance measure
\cite{note2}. The entanglement of distillation $E_D(\rho)$ is
defined as the maximal asymptotic ratio $M/N$ with respect to
all possible transformations which consist of LOCC \cite{Ra99}.
On the other hand, let us assume now that the parties possess $M$
two--qubit singlet states and they are able to transform them
into the state $\rho_M$ fulfilling
\be
\lim_{M\to\infty} D(\rho_M,\rho^{\otimes N}) \to 0,
\ee
for some integer $N$. The entanglement cost $E_C(\rho)$ is
defined as the minimal asymptotic ratio $M/N$ also with respect
to all LOCC \cite{Ha00}. The distillation process of a state
$\rho$ is irreversible if $E_D(\rho)<E_C(\rho)$.

Let us consider a density operator $\rho$ acting on $H_A\otimes
H_B$ and let us call $P$ the projector onto the range of $\rho$.
Then, we have that the entanglement cost of $\rho$ can be
bounded below as follows:

{\bf Theorem 1:} If $\langle e,f|P^{\otimes N}|e,f\rangle \le
\alpha^N$ for all normalized product vectors $|e,f\rangle\in (H_A)^{\otimes
N}\otimes (H_B)^{\otimes N}$ then $E_C(\rho)\ge -\log_2 \alpha$.

{\em Proof:} We will use the results of Ref.\ \cite{Ha00} where
it is shown that
\be
E_C(\rho)= \lim_{N\to\infty} \frac{E_f(\rho^{\otimes N})}{N},
\ee
where the limit exists. The entanglement of formation $E_f$ can
be determined by considering decompositions of the form
\cite{Be960,Wo98}
\be
\rho^{\otimes N}=\sum_i p_i |\Psi_i\rangle\langle \Psi_i|,
\ee
and minimizing the quantity $\sum_i p_i E(\Psi_i)$ with respect
to all possible decompositions. Here, $E$ denotes the entropy of
entanglement \cite{Be96}. Writing the Schmidt decomposition
$|\Psi_i\rangle=\sum_k c_{i,k} |e_{i,k},f_{i,k}\rangle$ we have
that
\be
|c_{i,k}|^2 \le \langle e_{i,k},f_{i,k}|P^{\otimes N}
|e_{i,k},f_{i,k}\rangle \le \alpha^N,
\ee
where the first inequality is a consequence of the fact that
$|\Psi_i\rangle\langle \Psi_i|\le P^{\otimes N}$, since all the
vectors $|\Psi_i\rangle$ must be in the range of $\rho^{\otimes
N}$. Given the fact that $\sum_k |c_{i,k}|^2=1$ we have that
$E(\Psi_i)\ge -N\log_2(\alpha)$ for all $i$ and therefore
$\sum_i p_i E(\Psi_i) \ge -N\log_2(\alpha)$ for all
decompositions. $\Box$

Let us consider the bound entangled state $\rho_b$ introduced in
Ref.\ \cite{Be99}, where $H_A=H_B=\C^3$. It is defined as
$\rho_b\equiv P_b/4$, where $P_b$ is a projector operator onto
the orthogonal complement to the subspace spanned by the
following vectors:
\bea
|0\rangle &\otimes& (|0\rangle +|1\rangle),\nonumber\\
(|0\rangle+|1\rangle) &\otimes& |2\rangle,\nonumber\\ |2\rangle
&\otimes& (|1\rangle +|2\rangle),\nonumber\\
(|1\rangle+|2\rangle)&\otimes& |0\rangle,\nonumber\\
(|0\rangle-|1\rangle+|2\rangle) &\otimes& (|0\rangle
-|1\rangle+|2\rangle).\nonumber
\eea
This state has a positive partial transposition, and therefore
it is not distillable ($E_D(\rho_b)=0$) \cite{Ho98}. Later on we will elaborate on this result. Our goal for the time being is
to show that $E_C(\rho_b)>0$.

We will use the following properties of the operator $P_b$:
\bma
\bea
\label{cond1}
\one + P_b &=& \sum_k |a_k,b_k\rangle\langle a_k,b_k|,\\
\label{cond2}
\alpha_1 &\equiv& \sup_{|e,f\rangle\ne 0} \langle
e,f|P_b|e,f\rangle <1.
\eea
\ema

The first equation indicates that the operator $\one+P_b$ is
separable. This can be proven by showing that the projector
operators $P_1\equiv\one-|a_0,a_0\rangle\langle a_0,a_0|$ and
$P_2\equiv P_b + |a_0,a_0\rangle\langle a_0,a_0|$ are both
separable, where $|a_0\rangle=(|0\rangle
-|1\rangle+|2\rangle)/\sqrt{3}$. First, choosing $|a_{1}\rangle$
and $|a_{2}\rangle$ in such a way that $\{|a_k\rangle\}_{k=0}^2$
forms an orthonormal basis one immediately sees that the range
of $P_1$ is spanned by the mutually orthogonal product vectors
$|a_{k_1},a_{k_2}\rangle$ where $k_1,k_2=0,1,2$ except for
$k_1=k_2=0$ and therefore $P_1$ is separable. Analogously, the
range of $P_2$ is spanned by the following mutually orthogonal
product vectors
\bea
|0\rangle &\otimes& (|0\rangle -|1\rangle),\nonumber\\
(|0\rangle-|1\rangle) &\otimes& |2\rangle,\nonumber\\ |2\rangle
&\otimes& (|1\rangle -|2\rangle),\nonumber\\
(|1\rangle-|2\rangle)&\otimes& |0\rangle,\nonumber\\
|1\rangle&\otimes& |1\rangle.\nonumber
\eea

The second equation is a direct consequence of the fact that the
range of $P_b$ contains no product vectors and that $\langle
e,f|P_b|e,f\rangle$ is a continuous function of $|e,f\rangle$
defined on a compact set so that it reaches its supremum
\cite{Te98}.

We will now show that for any normalized product vector
$|e^{N},f^{N}\rangle$ where $|e^{N}\rangle,|f^{N}\rangle \in
(\C^3)^{\otimes N}$,
\be
\label{cond3}
\langle e^{N},f^{N}|P_b^{\otimes N}|e^{N},f^{N}\rangle < \beta^N,
\ee
where $\beta\equiv(1+\alpha_1)/2< 1$. Then, the above theorem
readily implies that the entanglement cost $E_C$ of the bound
entangled state $\rho_{b} = P_b/4$ (and of any mixed state with
the same support $P_b$) is finite. We will use induction over
the number of copies $N$ to show that Eq. (\ref{cond3}) holds.
First, for $N=1$ it is true because of Eq. (\ref{cond2}) and
$\alpha_1<\beta$. Now, let us assume that it is true for a given
$N$. Then, for any product vector
$|e^{N\!+\!1},f^{N\!+\!1}\rangle \in (\C^3)^{\otimes N+1}\otimes
(\C^3)^{\otimes N+1}$ we have
\be
\label{cond4}
\langle e^{N\!+\!1},f^{N\!+\!1}|(\one+P)\!\otimes\!\!
\left[\one - \frac{1}{\beta^N}
P^{\otimes N}\right]\!|e^{N\!+\!1},f^{N\!+\!1}\rangle \geq 0.
\ee
The reason is that using (\ref{cond1}) and defining $|e_k
^{N}\rangle\equiv
\langle a_k|e^{N+1}\rangle$ and $|f_k^{N}\rangle\equiv
\langle b_k|f^{N+1}\rangle$ we can write the lhs of this equation as
\be
\sum_k \langle e^{N}_k,f^{N}_k|\left[\one -
\frac{1}{\beta^N} P^{\otimes N}\right]|e^{N}_k,f^{N}_k\rangle,
\ee
where all the terms in the sum are positive according to the
induction hypothesis (\ref{cond3}). Now, we can write
\be
(\one+P)\otimes \left[\one - \frac{1}{\beta^N} P^{\otimes
N}\right] \le \one + P\otimes\one - \frac{2}{\beta^N} P^{\otimes
N+1}.
\ee
Substituting this expression in Eq.\ (\ref{cond4}) we arrive at
\bea&&\langle e^{N+1},f^{N+1}|P^{\otimes N+1}|e^{N+1},f^{N+1}\rangle
\nonumber \\
&\leq& \beta^N \frac{1}{2}(1+\langle e^{N+1},f^{N+1}|P\otimes
\one|e^{N+1},f^{N+1}\rangle)\nonumber \\
&\le& \beta^{N+1},
\eea
as we wanted to prove.

The very same techniques can be applied to obtain lower bounds
for the entanglement cost $E_C$ also for more general mixed
states. Notice that the relevant ingredients we have used are
Eqs. (\ref{cond1}) and (\ref{cond2}), and that both conditions
are only concerned with the support of the state $\rho_b$.
Therefore, for any projector $P$ satisfying Eqs. (\ref{cond1})
and (\ref{cond2}), we get a non-trivial bound for the
entanglement cost of any state supported on it \cite{example}.

We move now to consider the distillability of the PPT state $\rho_b$ and the extension of the irreversibility result to catalytic (i.e. LOCCc-based) distillation. In \cite{Ho98} it was shown that inseparable states $\sigma_b$ with PPT can not be distilled into two-qubit singlet states $|\Psi\rangle$ using LOCC. The original proof relies on the fact that singlet states have a negative partial transposition (NPT), and LOCC cannot transform a PPT state into a NPT state.
 Notice, however, that if the parties initially share, in addition to the $N$ copies of the state $\sigma_b$, $L$ two-qubit singlet states $|\Psi\rangle$, then the original argument cannot be applied, because $\sigma_b^{\otimes N}\otimes |\Psi\rangle\langle \Psi|^{\otimes L}$ is a NPT state for any $L\geq 1$. And, then, maybe the transformation
\be
\sigma_b^{\otimes N}\otimes |\Psi\rangle\langle \Psi|^{\otimes L} \rightarrow  |\Psi\rangle\langle \Psi|^{\otimes M+L}
\label{disti}
\ee 
is asymptotically possible with some finite ratio $M/N$, in what would be a LOCCc distillation.  Thus, our previous results for the state $\rho_b$ do not yet exclude the possibility that in the large $N$ limit the equivalence
\be
\rho_b^{\otimes N}\otimes |\Psi\rangle\langle \Psi|^{\otimes L} \approx |\Psi\rangle\langle \Psi|^{\otimes M+L}
\label{LOCCq}
\ee
under LOCC holds, i.e., that the distillation of $\rho_b$ can be turned into a reversible process using entanglement catalysis.

The following, general result on bound entanglement readily implies that $\rho_b$ is not distillable even with LOCCc, thereby providing an example of asymptotic irreversibility also in a broader sense than that of LOCC.

{\bf Theorem 2:} Given $N$ copies of a PTT state $\sigma_b$ and $N'$ copies of some other state $\sigma$, the number of singlets that can be asymptotically distilled from them are, at most, the number of singlets required to create $\sigma^{\otimes N'}$:
\be
E_D(\sigma_b^{\otimes N}\otimes \sigma^{\otimes N'}) \leq N' E_C(\sigma),
\label{theorem2}
\ee
which in particular means that
\bea
E_D(\sigma_{b}\otimes \sigma) \leq E_C(\sigma); ~~~
E_D(\sigma_b)=0.
\eea
{\em Proof:} Consider the upper bound on distillability given by the logarithmic negativity $E_{\cal N}(\rho) \equiv \log_2 ||\rho^{T_B}||_1$ \cite{neg}, where $T_B$ stands for partial transposition and $||A||_1\equiv \mbox{tr} \sqrt{A^{\dagger}A}$ is the {\em trace norm} of $A$. $E_{\cal N}$ is an additive function which vanishes for PPT states and is $1$ for singlet states. Therefore, for general $N,L$ we have 
\be
E_{\cal N}(\sigma_b^{\otimes N}\otimes |\Psi\rangle\langle \Psi|^{\otimes L}) = L,
\label{bound}
\ee
 which implies that at most $L$ single states can be distilled from $\sigma_b^{\otimes N}\otimes |\Psi\rangle\langle \Psi|^{\otimes L}$. Setting $L\equiv N'E_C(\sigma)$, and observing that such number of singlets is sufficient to create $\sigma^{\otimes N'}$ locally, so that no more singlets can be distilled from $\sigma_b^{\otimes N}\otimes \sigma^{\otimes N'}$ than from $\sigma_b^{\otimes N}\otimes |\Psi\rangle\langle \Psi|^{\otimes L}$, we obtain Eq. (\ref{theorem2}). $\Box$

  Thus, an optimal LOCC transformation of the form of Eq. (\ref{disti}) has $M=L$, and pure entanglement does not help at distilling PPT states for LOCCc transformations \cite{LOCCq}. On the other hand it is easy to see that the entanglement cost of creating a mixed and a pure state is additive, so that \cite{comment}
\be
E_C(\sigma_b^{N}\otimes |\Psi\rangle\langle \Psi|^{\otimes L}) = N E_C(\rho) + L.
\label{addit}
\ee
 Then, when considering both the upper bound on distillability of Eq. (\ref{bound}) and the entanglement cost of Eq.(\ref{addit}), both applied to the PPT state $\rho_b$ for which we have proved that $E_C(\rho_b)>0$, we readily conclude that also the asymptotic LOCCc manipulation of entanglement is irreversible.

Summarizing, we have shown that the asymptotic entanglement cost
$E_C$ for locally preparing a given bound entangled state
$\rho_b$ is finite. Since no pure-state entanglement can be
distilled from the state $\rho_b$ even in the asymptotic limit 
[i.e. $E_D(\rho_b) = 0$], this result implies that the
asymptotic interconversion, by means of LOCC, between pure- and
mixed-state entanglement is in general not a reversible process. We have finally proved that such an irreversibility also occurs for asymptotic LOCCc transformations.

The authors acknowledge discussions with Jens Eisert,  Pawe\l  ~Horodecki, Maciej Lewenstein and Martin B. Plenio on the topic.   
This work was supported by the Austrian Science Foundation under
the SFB ``control and measurement of coherent quantum systems''
(Project 11), the European Community under the TMR network
ERB--FMRX--CT96--0087, project EQUIP (contract IST-1999-11053),
and contract HPMF-CT-1999-00200, the European Science
Foundation, and the Institute for Quantum Information GmbH. We
thank the Erwin Schr\"{o}dinger Institute for hosting us during the
quantum information program, where some of the ideas of this
work were developed.


\begin{references}
\bibitem{example}
 Consider, as an alternative example ---not using Eq. (\ref{cond1})--- of
how to get bounds for $E_C$, the projector $P$ spanned by the
vectors
\bea
\sqrt{\frac{2}{3}}|0,2\rangle + \sqrt{\frac{1}{3}}|1,0\rangle \nonumber\\
\sqrt{\frac{2}{3}}|2,0\rangle + \sqrt{\frac{1}{3}}|1,2\rangle \nonumber.
\eea
of a $\C^3\otimes \C^3$ system.
For any product vector $|e,f\rangle \in \C^3\otimes\C^3$ we have
that
\be
\langle e,f|P|e,f\rangle \leq \langle e|{\rm tr}_B P|e \rangle = \langle
e|\frac{2}{3}\one |e\rangle = \frac{2}{3},
\ee
and similarly, for any product vector $|e^{N},f^{N}\rangle$,
where $|e^{N}\rangle,|f^{N}\rangle \in (\C^3)^{\otimes N}$,
\be
\langle e^{N},f^{N}|P^{\otimes N}|e^{N},f^{N}\rangle \leq (2/3)^{N}.
\ee
Then the above theorem implies that the entanglement cost $E_C$
of any state $\rho$ supported on $P$ is at least $-\log_2
(2/3)$.
\bibitem{LOCCq} Let us marginally comment on a weaker result that also follows from Eq. (\ref{bound}). Namely, that PPT states cannot be distilled in the so-called LOCCq \cite{multiBennett} setting, when an asymptotically vanishing amount (per copy to be distilled) of pure entanglement is used in the distillation process. In a sense, this provides robustness to the original proof \cite{Ho98} of indistillability of PPT states.
\bibitem{comment} Since $E_C$ is defined in an asymptotic sense, it is also additive for identical copies of a state, and thus $E_C(\sigma_b^{\otimes N}) = NE_C(\sigma_b)$.
\end{references}
\end{document}